\documentclass[a4paper,twocolumn,3p,showkeys]{elsarticle} 
\usepackage{amsmath}  
\usepackage{amsfonts} 
\usepackage{graphicx} 
\usepackage{tikz}
\usepackage[siunitx]{circuitikz} 
\usetikzlibrary{shapes,arrows}
\usetikzlibrary{decorations.pathmorphing}
\usepackage{siunitx}
\usepackage{wasysym}
\usepackage{booktabs}
\usepackage[utf8]{inputenc}
\usepackage{subcaption}
\usepackage[]{lineno}
\usepackage[T1]{fontenc}
\usepackage{comment}

\makeatletter
\def\ps@pprintTitle{%
    \let\@oddhead\@empty
    \let\@evenhead\@empty
    \def\@oddfoot{\footnotesize\itshape
         {} \hfill\today}%
    \let\@evenfoot\@oddfoot
    }
\makeatother

\begin{document}

\title{Multispectral Photon-Counting for Medical Imaging and Beam Characterization}

\author[1]{E.~Brücken\corref{cor1}}
\ead{erik.brucken@iki.fi}
\cortext[cor1]{Corresponding author}
\author[1]{S.~Bharthuar}
\author[2]{M.~Emzir}
\author[1,3]{M.~Golovleva}
\author[1]{A.~G\"adda}
\author[2]{R.~Hostettler}
\author[6]{J.~Härkönen}
\author[1]{S.~Kirschenmann}
\author[1]{V.~Litichevskyi}
\author[1,3]{P.~Luukka}
\author[1]{L.~Martikainen}
\author[1]{T.~Naaranoja}
\author[1]{I.~Ninc\u{a}}
\author[1]{J.~Ott}
\author[3]{H.~Petrow}
\author[2]{Z.~Purisha}
\author[4]{T.~Siiskonen}
\author[2]{S.~S\"arkk\"a}
\author[4]{J.~Tikkanen}
\author[3]{T.~Tuuva}
\author[5]{A.~Winkler}
\address[1]{Helsinki Institute of Physics, Gustaf Hällströmin katu 2, FI-00014 University of Helsinki, Finland}
\address[2]{Aalto University, Otakaari 24, FI-00076 Aalto University, Finland}
\address[3]{Lappeenranta-Lahti University of Technology LUT, Yliopistonkatu 34, FI-53850 Lappeenranta, Finland}
\address[6]{Ru\dj er Bo\v{s}kovi\'c Institute, Zagreb 1000, Croatia}
\address[4]{Radiation and Nuclear Safety Authority (STUK), Laippatie 4, FI-00880 Helsinki, Finland}
\address[5]{Detection Technology Oyj, Otakaari 5A, FI-02150 Espoo, Finland}

\begin{abstract}
We present the current status of our project of developing a photon counting detector for medical imaging. An example motivation lays in producing a monitoring and dosimetry device for boron neutron capture therapy, currently not commercially available.  

Our approach combines in-house developed detectors based on cadmium telluride or thick silicon with readout chip technology developed for particle physics experiments at CERN.

Here we describe the manufacturing process of our sensors as well as the processing steps for the assembly of first prototypes. The prototypes use currently the PSI46digV2.1-r readout chip. The accompanying readout electronics chain that was used for first measurements will also be discussed. 
Finally we present an advanced algorithm developed by us for image reconstruction using such photon counting detectors with focus on boron neutron capture therapy. 

This work is conducted within a consortium of Finnish research groups from Helsinki Institute of Physics, Aalto University, Lappeenranta-Lahti University of Technology LUT and Radiation and Nuclear Safety Authority (STUK) under the RADDESS program of Academy of Finland.
\end{abstract}

\begin{keyword}Solid state detectors, X-ray detectors, Gamma detectors, Neutron detectors, Instrumentation for hadron therapy, Medical-image reconstruction methods and algorithms.
\end{keyword}

\maketitle

\section{Introduction}
\label{intro}

Next generation detection systems operating in multispectral mode have the potential to improve diagnostic capabilities in medical imaging substantially in terms of efficiency, image quality and lower patient dose (see~e.g.\cite{PCDCT,mohamed}). 
Our approach utilizes direct conversion radiation detectors operating in Photon Counting (PC) mode. 
Our strategy is to use several sensor materials including thick silicon, high-Z semiconductor materials (CdTe/CdZnTe) and silicon enhanced by scintillator (SiS) material together with pixel Read-Out Chips (ROC) running in PC mode. Due to our involvement in high energy physics, in particular in the CMS Tracker at CERN, we have access to existing solutions of ROCs that are capable of working in the PC mode.

\begin{figure*}[!htb]
\centering 
\hspace{-1em}\includegraphics[width=.8\textwidth]{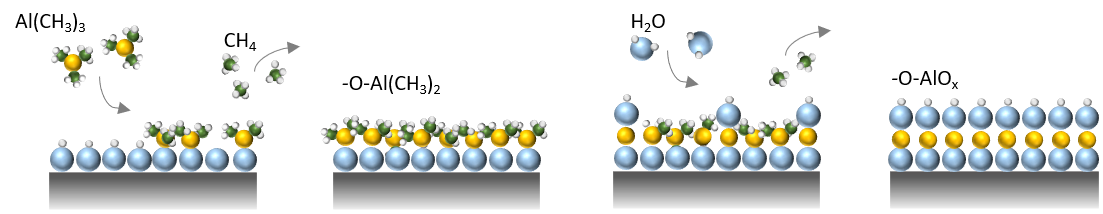}
\caption{\label{fig:jenni} Visualization of one ALD cycle to produce thin Al$_2$O$_3$ films on CdTe.}
\end{figure*}
The main focus lies on the utilization of CdTe. Due to its high quantum efficiency (high Z) it outperforms silicon in terms of photon radiation absorption.
The band gap of 1.44 keV is large enough to allow operation at room temperatures without significant thermal noise, but small enough to achieve good energy resolution~\cite{delsordo, bertuccio}.
However, CdTe crystals are, at present times, difficult to grow and are only available in small form-factors containing a variety of defects. Therefore, we apply a thorough quality assurance that enables us to choose the best crystals for detector fabrication.

After manufacturing first successful prototypes for the proof-of-concept, we are now focusing on the processing of the CdTe crystals and thick Si wafers at Micronova Nanofabrication Centre in Espoo, Finland. Processed sensors will then be flip-chip bonded with the ROCs, which is a critical step in the detector production. Due to the intrinsic material properties of CdTe, bump bonding has to be done at lower temperatures of <150\,$^{\circ}$C compared to silicon sensors, thus usual standard materials cannot be used. A feasible approach is to employ bumps made of Indium that allow bonding at low temperatures.

In addition to detector development, other crucial tasks related to this project are: the evolution from single module to detector arrays and its electronic readout; the advanced data analysis and image reconstruction; and prototype validation to guarantee repeatability and long term stability.

\section{Detector manufacturing}
We use detector grade cadmium telluride crystals with (111) orientation from Acrorad Ltd.\,in sizes of \mbox{$10 \times10 \times 1$~mm$^3$}. The resistivity is above $10^9\,\Omega$cm.

The detector is of Schottky type with a design that is currently matched to the layout of the PSI46digV2.1-r ROC. The ROC has 4160 pixels organized in 52 columns and 80 rows with an active area of $8\times 7.6$~mm$^2$. The individual pixel size is $150 \times 100\,\mu$m$^2$. 

The entire processing of the CdTe crystals, described in detail in \cite{akiko_processing}, is done in Micronova, the Centre for Micro and Nanotechnology in Finland. The processing starts with depositing a thin layer of Al$_2$O$_3$ as field insulation to the bare crystal via atomic layer deposition (ALD) method. ALD consists of cycles of self-terminating gas-solid reactions that allow the growth of accurate thin films on substrates.  An illustration of one cycle for the growth of Al$_2$O$_3$ is shown in Figure~\ref{fig:jenni}. 
The growth takes place in a Beneq TFS-500 reactor at rather low temperatures of~120$^{\circ}$\,C. First, a pulse of trimethylaluminium (Al(CH\textsubscript{3})\textsubscript{3}) as metal precursor is injected followed by a purging pulse of N\textsubscript{2}. Next, a H\textsubscript{2}O pulse is injected, again followed by a N\textsubscript{2} purge. This cycle is repeated until the desired layer thickness -- here around 90\,nm -- is achieved. 

As next steps alignment marks of titanium-tungsten (TiW) and opening contacts to the passivation layer via wet chemical etching are formed. Contact metallization of TiW and gold (Au) is sputtered and finalized via lift-off process. The backside of the pixel sensor is processed similar to the front side with sputtered TiW as contacts. This results in a single electrode design on the backside with local openings to the CdTe. Finally, the front side is refined with electro-less Ni growth and Au metallization as under bump metallization (UBM).

The patterned structure on the backside is based on ideas of Kramberger and Contarato \cite{krami}.  It is expected that such electrode design restricts the weighting fields of the electrodes and thus should improve the charge collection efficiency and energy resolution of the detector.

\section{Quality assurance of the raw material}

It is  well known that growing detector grade CdTe crystals is very challenging. Currently, maximally 2 inch ingots are available and even those not in entirely mono-crystalline form. They also show a variety of crystallographic defects such as grain and twin boundaries, fractures, tellurium inclusions of different forms. All those affect the detector performance substantially.
\begin{figure*}[!htb]
\centering
\begin{subfigure}{.48\textwidth}
  \centering
\includegraphics[height=3.9cm]{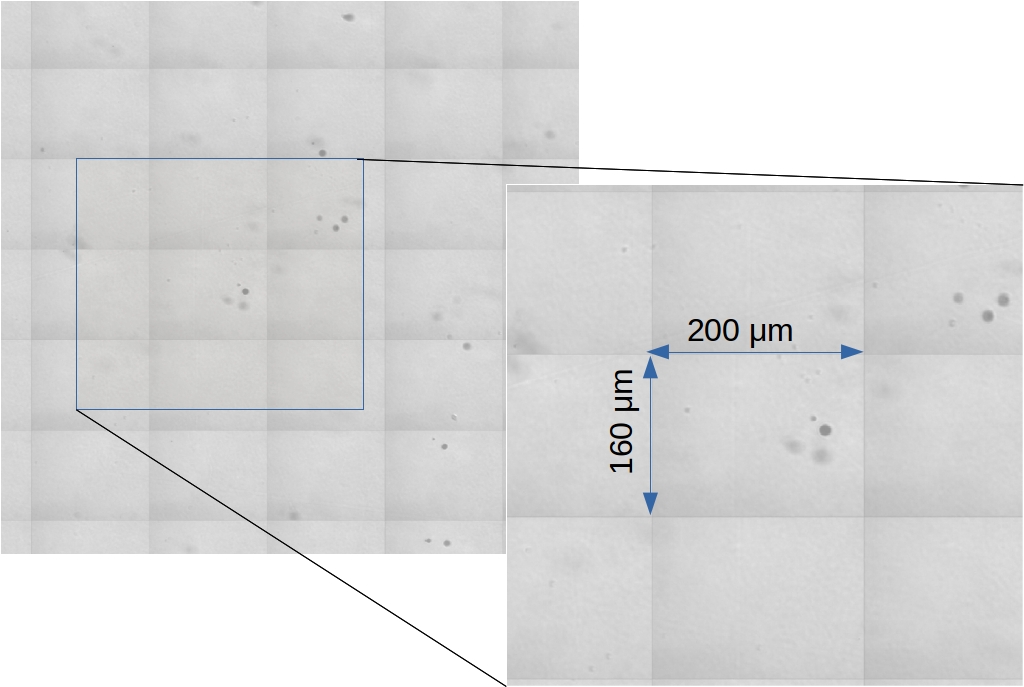}
  \caption{}
  \label{fig:IRimg}
\end{subfigure}
\quad
\begin{subfigure}{.48\textwidth}
  \centering
\includegraphics[height=3.9cm]{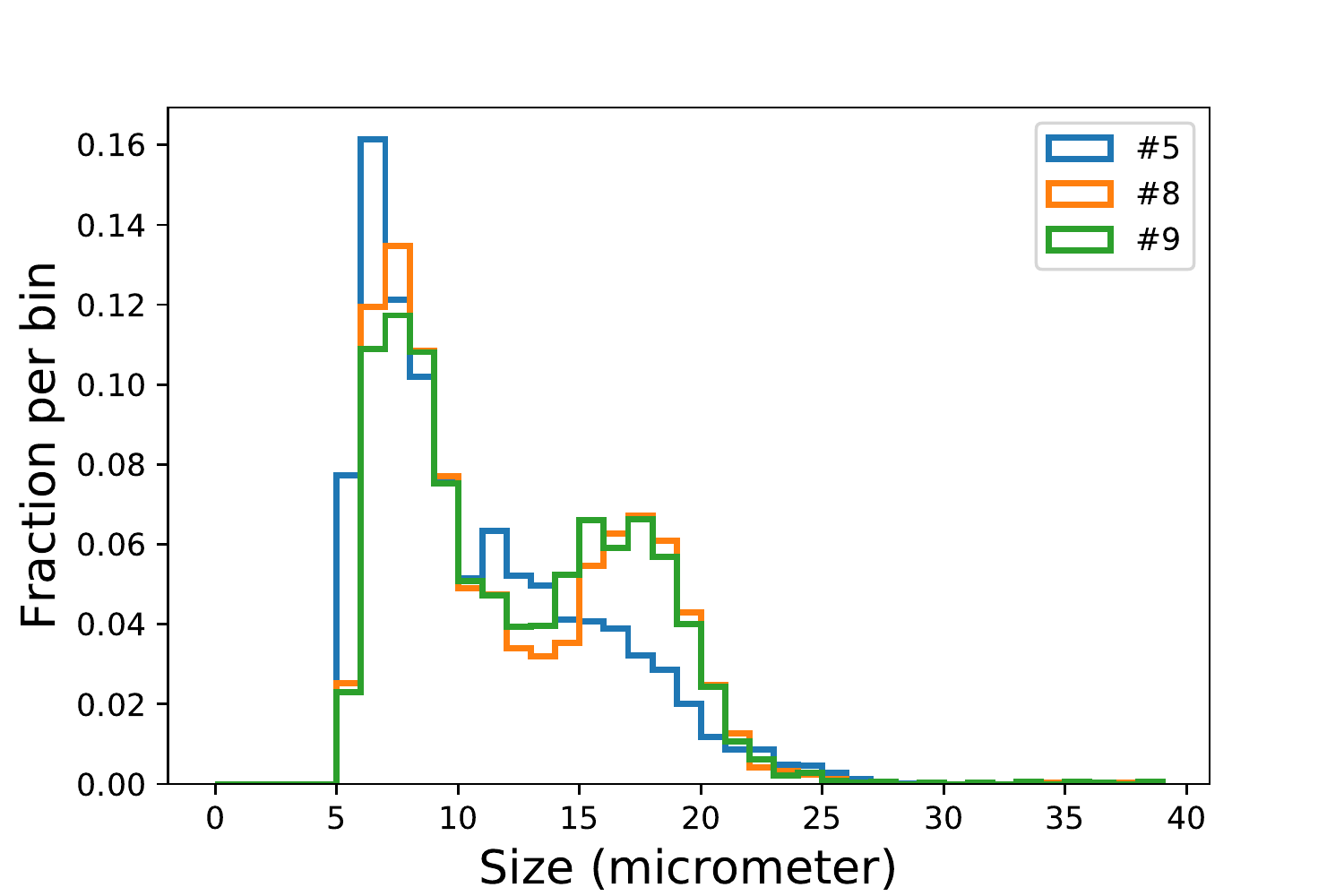}
  \caption{}
  \label{fig:IRhisto}
\end{subfigure}
\caption{\label{fig:stefanie} (\ref{fig:IRimg}) Example IR images from a scanned CdTe crystal showing tellurium inclusions, (\ref{fig:IRhisto}) size distribution of tellurium inclusions of scanned CdTe crystals.}
\end{figure*}

Therefore, a thorough quality assurance is done before starting the processing of the raw material. We use a commercially available infra-red (IR) spectrometer that we modified into a scanning IR imaging device. As CdTe crystals are transparent for IR light, we are able to scan the whole crystal for defects \cite{winkiIR}. Tellurium inclusions are not transparent to IR light, thus we can identify those down to sizes of a few $\mu$m, close to the diffraction limit. An example excerpt of an IR image of one CdTe crystal is shown in Figure~\ref{fig:IRimg}. 
As outcome we receive detailed 3D distributions and maps of tellurium inclusions and can classify those according to shape and size. Example histograms showing the size distributions of tellurium inclusions of three CdTe crystals are shown in Figure\,\ref{fig:IRhisto}.

Currently, we study in collaboration with the Ru\dj er Bo\v{s}kovi\'c Institute (RBI) how the tellurium inclusions affect the local charge collection efficiency (CCE)~\cite{matti,matti2}. After thorough IR scanning of the bare crystals, simple Schottky type diode structures were produced at Micronova. The CdTe diodes were then studied in detail using scanning micro proton beam (IBIC) and scanning transient current technique (TCT) at RBI.
In case of IBIC, protons of 10\,MeV are injected locally into the CdTe diode. Most charge is produced at about 40\,$\mu$m below the surface (Bragg peak). The resulting signal of the biased diode is measured with charge sensitive data acquisition (DAQ). 
The investigation is currently ongoing to proof and quantify the direct correlation of defects seen in the IR imaging and the CCE drop in the IBIC and TCT measurements.

\section{Readout electronics}
\begin{figure*}[]
\centering
\begin{subfigure}{.4\textwidth}
  \centering
\includegraphics[height=3.1cm]{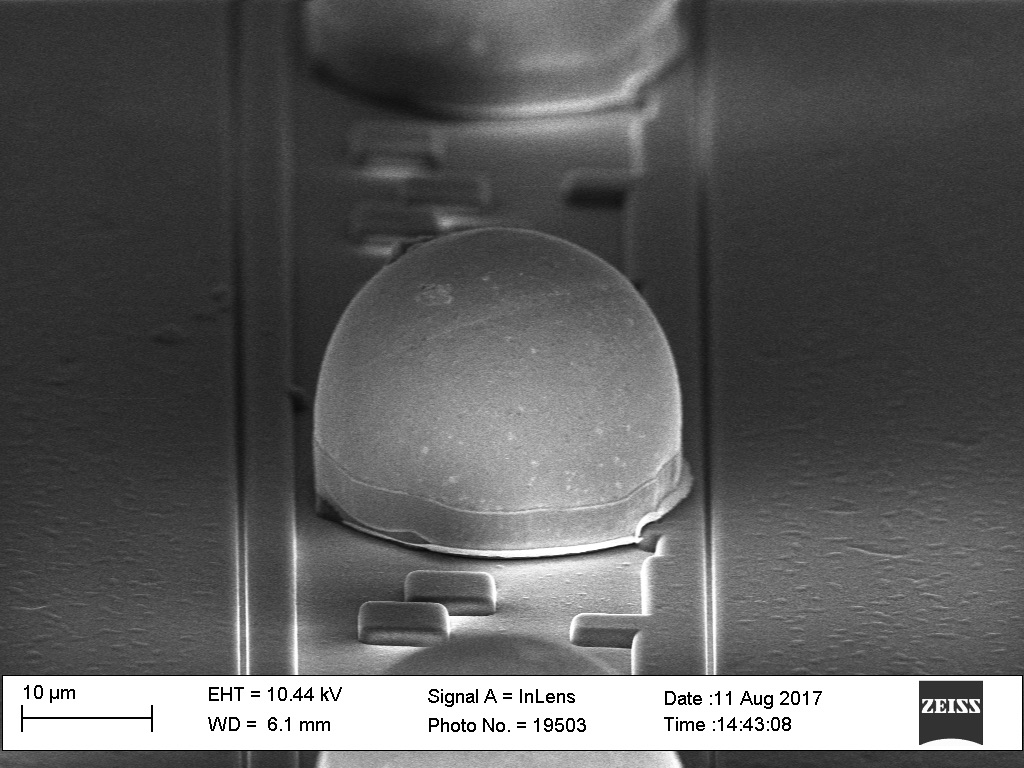}
  \caption{}
  \label{fig:indium}
\end{subfigure}
\quad
\begin{subfigure}{.5\textwidth}
  \centering
\includegraphics[height=3.1cm]{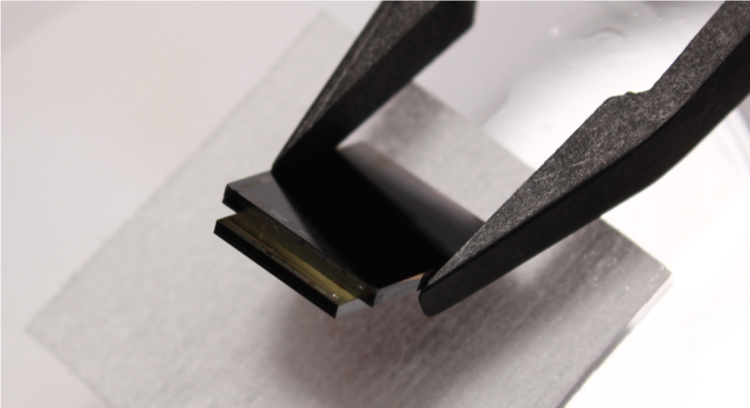}
  \caption{}
  \label{fig:prototype}
\end{subfigure}
\caption{\label{fig:detector} (\ref{fig:indium}) Microscopic image of part of the PSI46digV2.1-r readout chip with Indium bumps on the contact pads, (\ref{fig:prototype}) a manufactured prototype detector.}
\end{figure*}

The readout chip (ROC) used was developed by the Paul Scherrer Institute (PSI) for the CMS phase 1 pixel detector \cite{cmspixel,beat}. It is a radiation hard (>25\,kGy) CMOS ASIC fabricated in 250~nm technology by IBM. The ROC is photon counting capable with pre-amplification, shaping and digitisation stage for each pixel. The signal pulse hight is converted via an 8-bit ADC. The charge threshold lies around 1.5\,ke$^-$ with a resolution of approximately 120\,e$^-$. The ROC wafers used were populated with Indium bumps, suitable for low temperature flip chip bonding as needed for CdTe based sensors. A microscopic image of part of the PSI46digV2.1-r ROC with Indium bumps on the contact pads is shown in Figure~\ref{fig:indium}. A ready, bump bonded detector is shown in Figure~\ref{fig:prototype}.

The detector prototypes are wire bonded to a front end card connected to the Detector Test Board (DTB) that was developed by PSI for testing the ROCs of the PSI46 type family. The DTB features an Altera FPGA and can be interfaced with a PC via USB2.0 or Gigabit Ethernet connection.

\begin{figure}[!hb]
\centering 
\includegraphics[width=.45\textwidth]{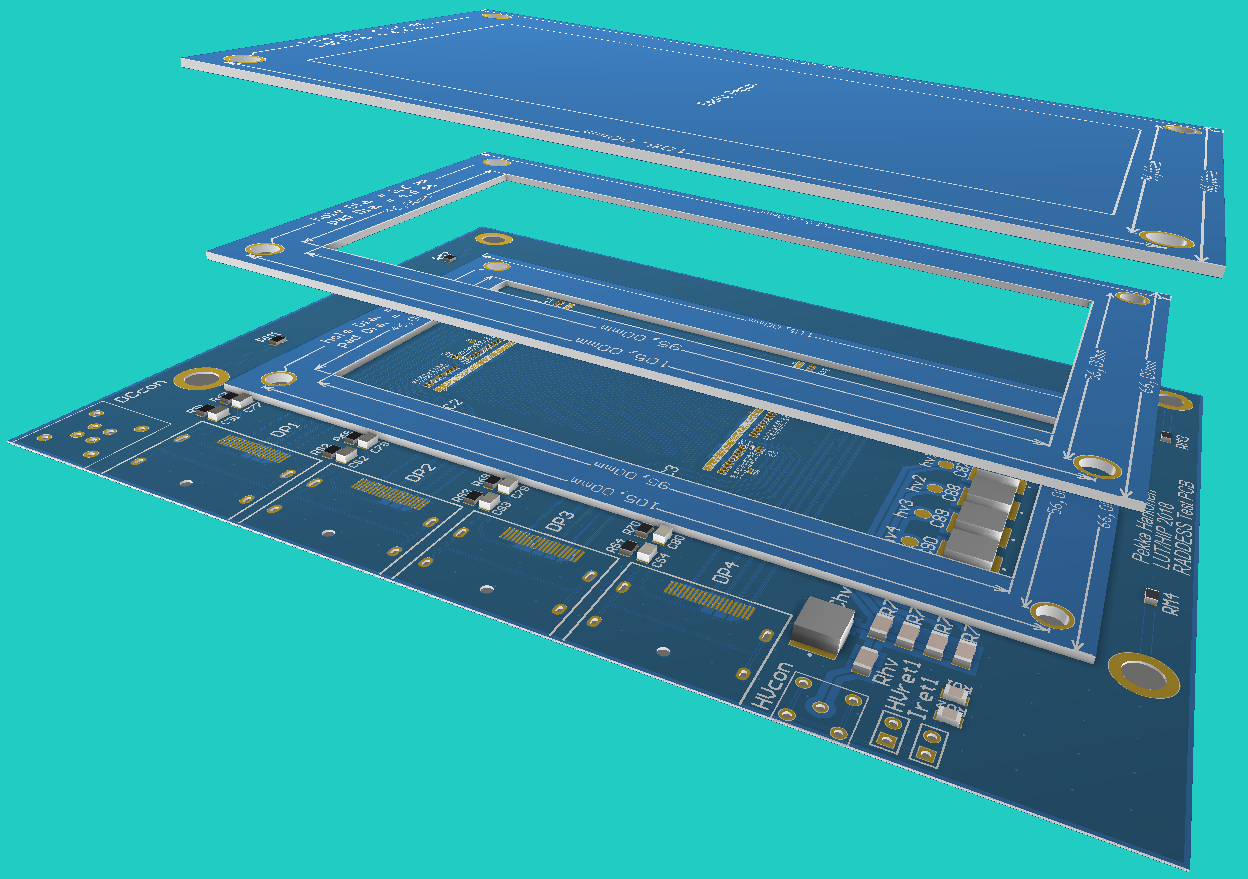}
\caption{\label{fig:petrow} The design of the high density interconnect for hosting up to 4 detectors based on the RD53a ROC design.} 
\end{figure}

In parallel of testing the existing PSI46 ROC infrastructure we started working on hardware to utilize the soon to be available RD53a ROC. This ROC is designed and constructed within the CERN RD53 collaboration for the phase 2 upgrade of the tracking detectors for the CMS and ATLAS experiment at the Large Hadron Collider~\cite{rd53}. In Figure~\ref{fig:petrow}, a rendering of the high density interconnect (HDI) is shown that we design to host up to four hybrid detectors based on the RD53 design.
The HDI is fitted with individual communication and data lines for each ROC and includes separate high voltage biasing for the detectors. The communication line to a FPGA board will be connected via FMC-DisplayPort adapter. In the elevated view in Figure~\ref{fig:petrow}, possible supplemental cover frames are shown, that can be stacked on top of the HDI, e.g. for transport protection.

\section{Test results}

\begin{figure*}[]
\centering 
\begin{subfigure}{.32\textwidth}
  \centering
  \includegraphics[width=\textwidth]{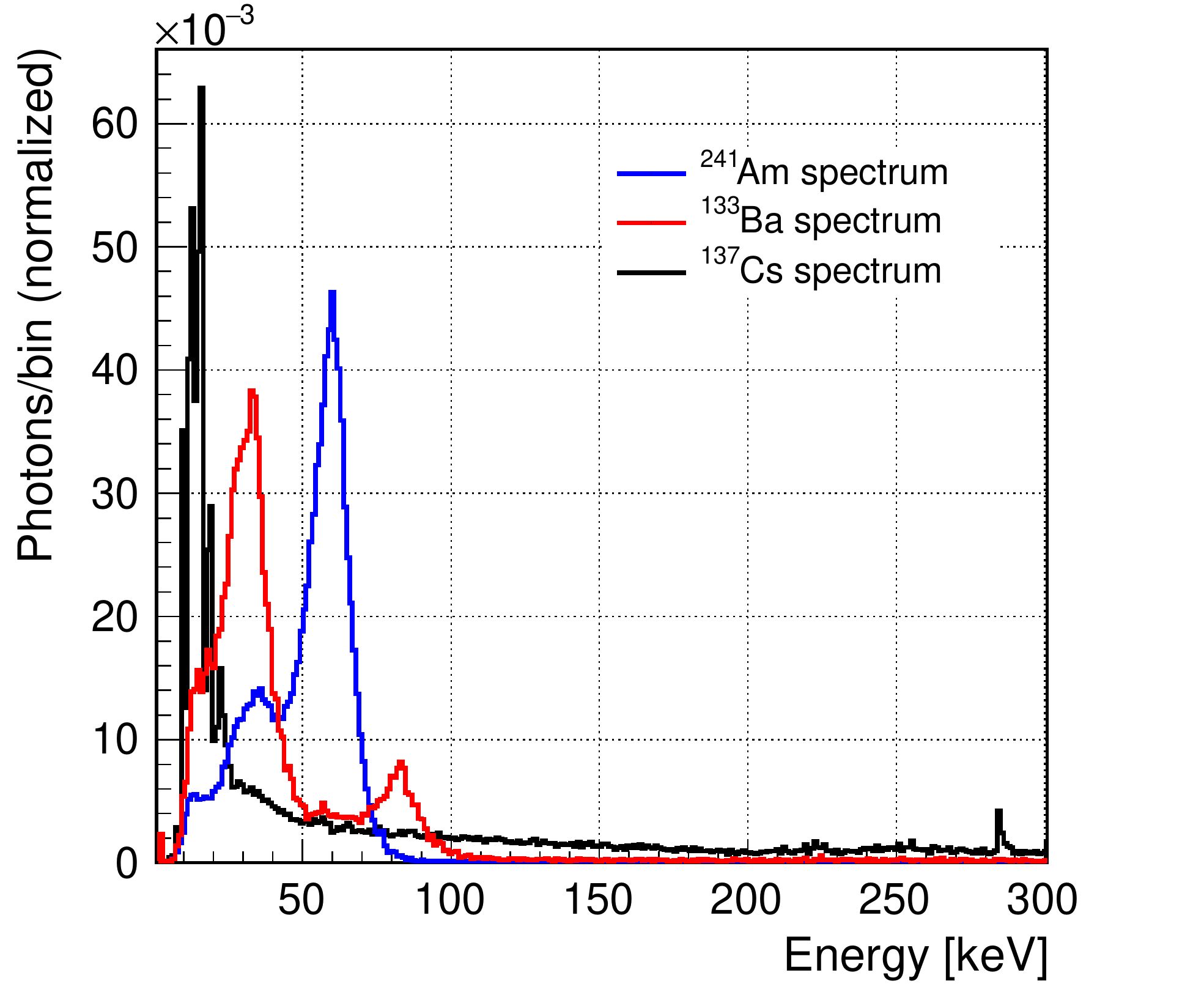}
  \caption{}
  \label{fig:spectra_all}
\end{subfigure}
\,
\begin{subfigure}{.32\textwidth}
  \centering
  \includegraphics[width=\textwidth]{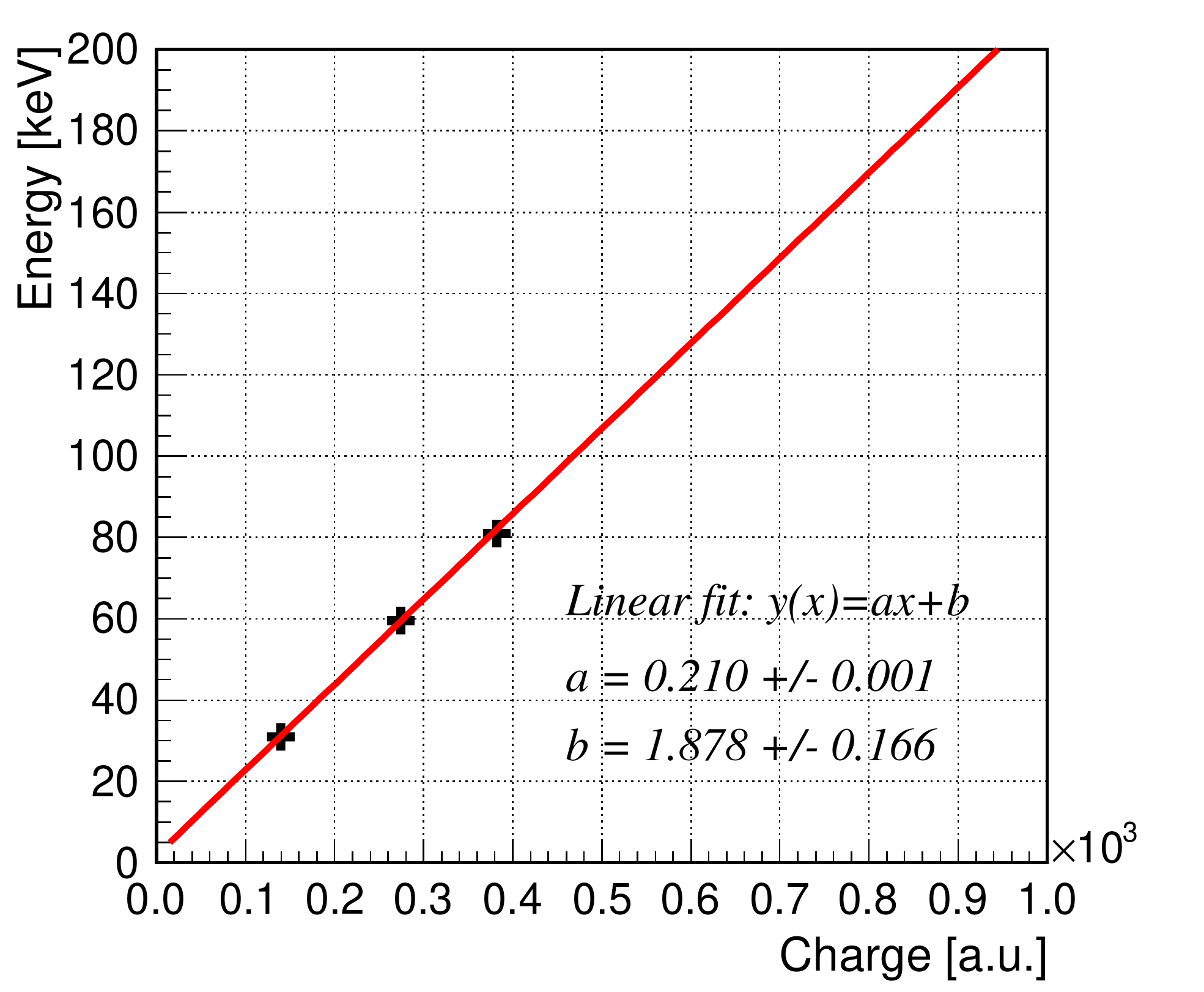}
  \caption{}
  \label{fig:calibration}
\end{subfigure}
\,
\begin{subfigure}{.32\textwidth}
  \centering
  \includegraphics[width=\textwidth]{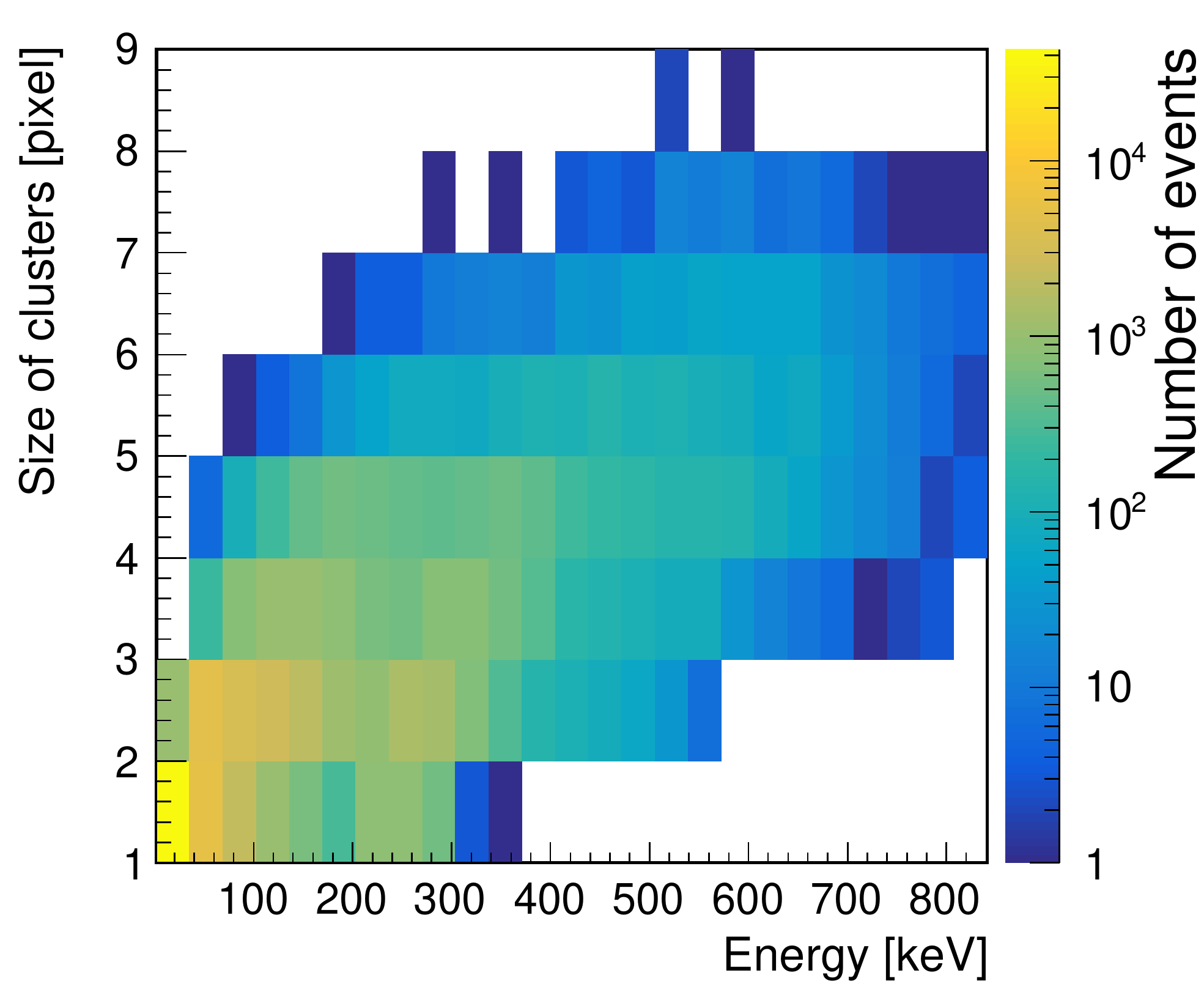}
  \caption{}
  \label{fig:chargesharing}
\end{subfigure}
\caption{\label{fig:cdteresults} Preliminary results of the irradiation of the CdTe pixel detector prototype. (\ref{fig:spectra_all}) Spectra of calibration sources, (\ref{fig:calibration}) energy calibration using characteristic emissions of the calibration sources and (\ref{fig:chargesharing}) the pixel cluster size of the reconstructed single photon events versus the energy. }
\end{figure*}

Preliminary results of the irradiation of our prototype are shown in Figure~\ref{fig:cdteresults}. The detector was biased with~-250\,V from the backside with a Keithley picoammeter/voltage source 6487 via high voltage filter. A current of 10\,$\mu$A was measured. As calibration sources we used~$^{241}$Am, $^{133}$Ba and~$^{137}$Cs. The recorded spectra are shown in Figure~\ref{fig:spectra_all}. The characteristic gamma-ray peak of~$^{241}$Am at 59.5\,keV and the X- and gamma-ray peaks of~$^{133}$Ba at 31\,keV and 81\,keV respectively, were used for energy calibration show in Figure~\ref{fig:calibration}. 
As the activity of our~$^{137}$Cs source was rather low and the detector is currently not completely optimized, we still get a suboptimal energy resolution (27\% @ 59.5\,keV of~$^{241}$Am). Thus, gaining high statistics and resolving the characteristic 662\,keV gamma-ray peak of the~$^{137}$Cs source, was not possible. Figure~\ref{fig:chargesharing} shows the reconstructed pixel cluster size versus the recorded photon energies. 

The peak in the~$^{137}$Cs spectrum close to 300\,keV, clearly visible in Figure~\ref{fig:spectra_all}, refers to the saturation limit in the readout electronics of single pixels.
This saturation limit will not allow for the direct measurement of the full charge of higher energetic gamma-rays within one pixel. However, we believe that after fine tuning the ROC and taking into account charge-sharing over neighbouring pixels by clusterization, we will be able to resolve high energetic gamma-rays up to 600\,keV. This would be the energy range of interest for an application within BNCT (see section \ref{BNCT} below). 
However, in future we plan to utilize a readout solution with higher dynamic range to avoid saturation effects.

\section{Application for BNCT including novel algorithms for image processing}
\label{BNCT}

A possible application of our planned detector array is within the boron neutron capture therapy  (BNCT). This is a binary therapy using an epi-thermal neutron beam (E$_{\mathrm{neutron}}\,\approx\,10$\,keV) and is currently targeted at inoperable malignant cancers, in particularly within the head and neck region \cite{Sweet1951}.

For BNCT, $^{10}$B is added to a tumor-targeting drug as part of the molecule/compound. The drug is then administered to the patient, followed by an irradiation of the cancerous area with the epi-thermal neutron beam. These neutrons get captured by $^{10}$B and the nuclear reaction: $$^{10}\mathsf{B}+\mathsf{n} \rightarrow \mathsf{^{11}B}^* \rightarrow \mathsf{^{7}Li^{\ast}} +\alpha \rightarrow \mathsf{^{7}Li} +\gamma$$
occurs. The therapeutical effective process is induced by the linear energy transfer of the $\alpha$-particle within the radius of one cell, which will effectively destroy the cancerous cell.

A key requirement of BNCT is $^{10}$B dosimetry, i.e.\,the determination of the $^{10}$B concentration ratio in the cancerous to healthy tissues, as well as monitoring this concentration ratio during the treatment. For a successful irradiation session, a ratio of~3.5:1 of the tumours versus healthy tissue is essential \cite{Kankaanranta:2012dp}. The determination of this ratio within a patient is currently based on blood measurements from patients prior to their treatments and extensive computational simulations.

PC detectors as proposed in this study, could provide this ratio in real time and during treatment. Furthermore, a BNCT-SPECT (BNCT- single photon emission computed tomography) like system based on PC detectors could further provide the spatial distribution of the~$^{10}$B within the patient. This allows to image the tumour irradiation in three dimension and calculate or correct the predetermined patient dose based on the extracted $^{10}$B concentration.
\begin{figure*}[!htb]
\centering 
\begin{subfigure}{.19\textwidth}
  \centering
  \includegraphics[width=\textwidth]{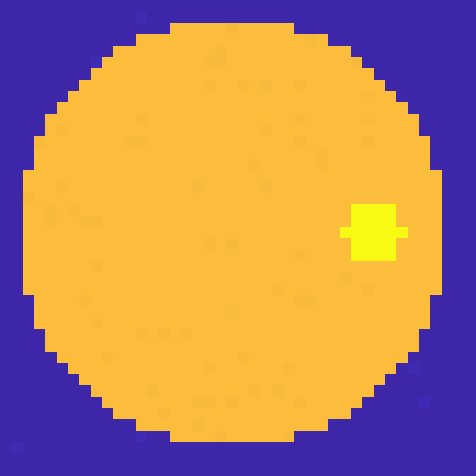}
  \caption{}
  \label{fig:object}
\end{subfigure}
\,
\begin{subfigure}{.19\textwidth}
  \centering
  \includegraphics[width=\textwidth]{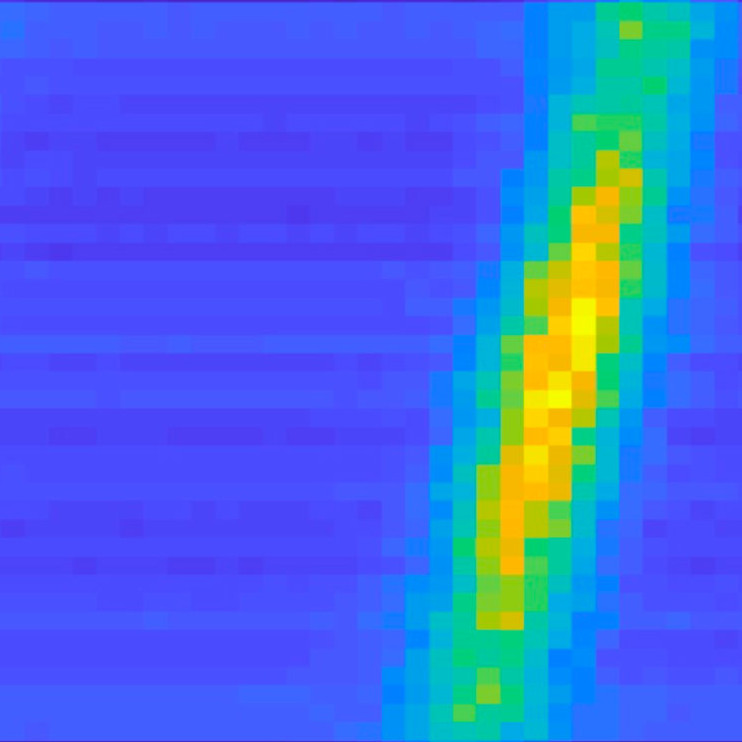}
  \caption{}
  \label{fig:sinogram}
\end{subfigure}
\,
\begin{subfigure}{.19\textwidth}
  \centering
  \includegraphics[width=\textwidth]{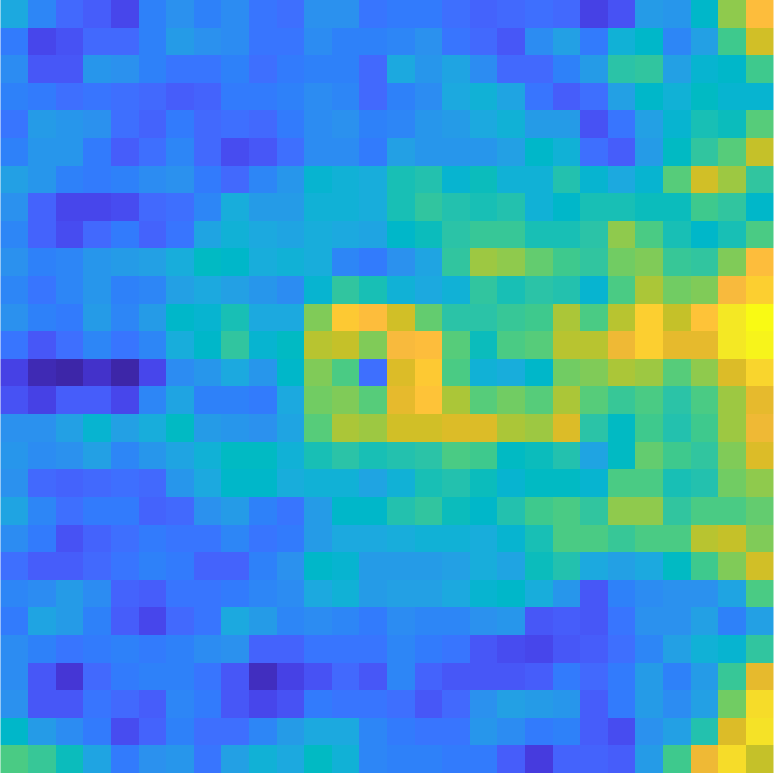}
  \caption{}
  \label{fig:fbp}
\end{subfigure}
\,
\begin{subfigure}{.19\textwidth}
  \centering
  \includegraphics[width=\textwidth]{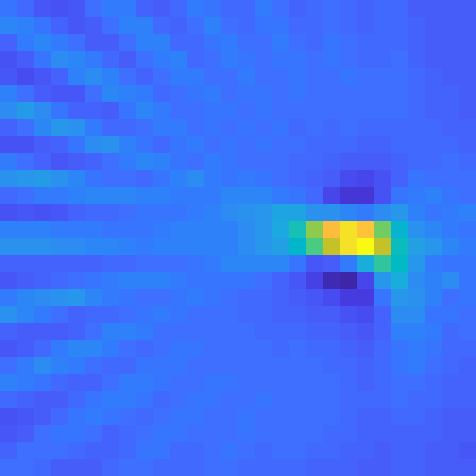}
  \caption{}
  \label{fig:gaussianprior}\end{subfigure}
\caption{\label{fig:advancedalgorithm} Simulation results illustrating the capabilities of the developed reconstruction algorithm. (\ref{fig:object}) object with enclosed boron, (\ref{fig:sinogram}) the data sets (sinogram) formed by the posterior distribution (eq.\,(\ref{eq:posterior})), (\ref{fig:fbp}) standard filtered back projection method and (\ref{fig:gaussianprior}) the FBP reconstruction from data sets (eq.\,(\ref{eq:posterior})). }
\end{figure*}

Such a detector system that enables treatment monitoring, is one of the last missing pieces for BNCT to become an accepted alternative radiation therapy \cite{Barth2012}. BNCT-SPECT imaging can be done using the prompt gamma photons from the relaxation of the $^{7}$Li$^{\ast}$ atom at~478\,keV, or by measuring the scattered neutrons that reach the detector~\cite{winki}. If the target is to measure the~478\,keV photons directly, then the main challenge is to extract this signal from the measured background, i.e., the neutron induced background. Alternatively, this neutron background can be defined as a signal, which is inversely dependent on the signal coming from the boron neutron reaction, as shown in~\cite{winki}. In this case the scattered neutrons can be detected directly by neutron capture of cadmium ($^{113}$Cd) that is part of the CdTe, or CdZnTe sensor, through the nuclear reaction: 

$$^{113}\mathsf{Cd}+\mathsf{n} \rightarrow ^{114}\mathsf{Cd^{\ast}} \rightarrow ^{114}\mathsf{Cd} +\gamma$$

The accompanying prompt gamma photon from the de-relaxation process of the $^{114}$Cd$^{\ast}$ has an energy of~558\,keV. However, this signal is substantially larger, as more neutrons scatter towards the detector, then the boron neutron capture reactions occur. Thus, it is easier to measure the neutron signal.

Despite the technical challenges of constructing a PC detector that has a sufficiently high energy resolution to distinguish between the~478\,keV and the~558\,keV prompt gammas, another practical problem arises to determine and image the $^{10}$B distribution within the patient.

In order to perform a SPECT like image reconstruction and measure the $^{10}$B dose, anti-scatter grids (ASG) are needed to increase image contrast. Otherwise the~478\,keV photons originating from the boron neutron capture (BNC) reaction cannot be distinguished from the strong gamma and neutron background present during the therapy \cite{Kobayashi:2000ip, Minsky:2011ez}. This requires heavy collimators, which reduce the BNC signal further and induce additional patient dose through generation of secondary radiation that is generated within the collimator structure. In case of neutron imaging, no feasible solution exists to realise an anti-scatter grid without increasing the patient dose as well.

One approach to solve this problem is to transfer the contrast enhancement from the physical ASG to the off-line image reconstruction algorithms. 

Let us suppose that we have successfully measured scattered neutrons from the irradiation of the sample with our PC detector array. In this particular case, only limited angle measurement data is recorded in the detector \cite{Frikel2013}

We are interested in estimating the radiotracer activity~$\mathbf x^b$ corresponding to the concentration of Boron inside the sample \cite{Jaszczak1980,Wernick2004}. Let us denote the measured data for each pixel in the detector array as the vector $\mathbf y$. The effect of scattering in the process can be accounted for by modelling the measurements $\mathbf y$ as noisy function of an ideal measurement $\mathbf y^b$ that would be obtained if there was no scattering \cite[\textsection 5.4]{NationalResearchCouncils1996}.

There are many ways to model the scattering process \cite{Conti2012,Hutton2011}, however, at this stage we use the following simple statistical model
\begin{equation}
    p(\mathbf{y} \mid \mathbf{y^b}) = \mathcal{N}(\mathbf{y} \mid \mathbf{S} \mathbf{y^b}, \mathbf{C}), \label{eq:y_cond_yb}
\end{equation}
where $\mathbf{S}$ is a linear operator that approximates the scattering process, and $\mathbf{C}$ is the covariance of a Gaussian noise process modelling the stochastic effects and noises. The ideal measurement can be modelled as follows:
\begin{equation}
\mathbf y^b = {\mbox{I}_0} \mathbf{A} \mathbf x^b,
\end{equation} 
where ${\mbox{I}_0}$ is the total number of measured events (e.g. photons, or neutrons), $\mathbf{A}$ the noise-free tomographic measurement matrix,  and $\mathbf x^b$ is the boron density. In our approach, the scattering operator~$\mathbf{S}$ is estimated empirically from the data, and it is varies from a measurement type to another.

Given the scattering model, the posterior distribution for $\mathbf y^b$ has the following form:
\begin{equation}p({{\mathbf y}}^b\mid {\mathbf y}) \propto p({\mathbf y}\mid {{\mathbf y}}^b)p({{\mathbf y}^b}),\label{eq:posterior}
\end{equation}
where $p({{\mathbf y}^b)}$ is prior distribution for the signal. After obtaining an estimate of $\mathbf y^b$ from above we can then use the standard Filtered Back Projection (FBP) method to reconstruct the boron density $\mathbf x^b$. Although this model has limitations, its advantage is that the reconstruction algorithm remains very fast as an explicitly inverse problems solver \cite{Tarantola2005,Purisha2019} is not needed. However, the accuracy of the approximation might be limited.

Results are presented in Figure~\ref{fig:advancedalgorithm}. 
On the left in Figure~\ref{fig:object} the modelled phantom with boron inclusion is shown. 
The data sets formed by the posterior distribution (eq.\,(\ref{eq:posterior})) is shown in Figure~\ref{fig:sinogram} and the result of applying the standard Filtered Back Projection (FBP) method is shown in Figure~\ref{fig:fbp}. The result of applying FBP method to the data (Figure~\ref{fig:sinogram})  is shown in Figure~\ref{fig:gaussianprior}. Without doubts one can see the superior performance of the probabilistic approach combining with FBP compared to standard FBP in case no anti-scatter grid is available. This image reconstruction method is expected to work for both, neutron and photon based data.

\section{Conclusions and outlook}

We have successfully constructed prototypes of a PC pixel detector based on CdTe sensors and the \mbox{PSI46digV2.1-r} ROC. First results show that we can successfully resolve prominent photo-peaks from the characteristic emissions of the calibration sources~$^{241}$Am and~$^{133}$Ba. However, further studies of the detector and fine adjustments of the DAQ are needed to achieve good acceptance and efficiencies for the desired energy range up to~600\,keV and sufficient energy resolution. It is planned to also fabricate CdTe based sensors of~2\,mm thickness for better absorption efficiency for high energetic gamma photons. 

As the development of the SPECT image reconstruction is already advanced, our emphasis is to prepare a test for our detectors at the local BNCT facility of the Helsinki University hospital, that is currently under commissioning.

\section*{Acknowledgements}

We acknowledge the funding by Academy of Finland project, number 314473, \emph{Multispectral photon-counting for medical imaging and beam characterization}.
The sample fabrication was performed at Micronova Nanofabrication Centre in Espoo, Finland within the OtaNano research infrastructure at Aalto University.
We also thank the staff of the joint detector laboratory of the University of Helsinki and the Helsinki Institute of Physics for their support.

\end{document}